\documentclass{aa}
\usepackage{graphics}

\begin{document}

\thesaurus{10 
           (08.02.7; 08.08.1; 08.16.3; 10.07.3 NGC 6229; 10.07.3 M\,5)}

\title{The outer-halo globular cluster NGC~6229.\\III. Deep CCD photometry}

\author{
       J.~Borissova\inst{1}
       \and
       M.~Catelan\inst{2}\thanks{Hubble Fellow.}\fnmsep\thanks{Visiting
                     Scientist, Universities Space Research Association.}
       \and
       F.~R.~Ferraro\inst{3}
       \and
       N.~Spassova\inst{1}
       \and
       R.~Buonanno\inst{4}
       \and
       G.~Iannicola\inst{4}
       \and
       T.~Richtler\inst{5}\thanks{Visiting Astronomer, German-Spanish 
                     Astronomical Center, Calar Alto, operated by the 
                     Max~Planck-Institut f\"ur Astronomie jointly with 
                     the Spanish National Commission for Astronomy.}
       \and
       A.~V.~Sweigart\inst{2}
}

\offprints{J.~Borissova}

\institute {Institute of Astronomy, Bulgarian Academy of Sciences,
72~Tsarigradsko chauss\`ee, BG\,--\,1784 Sofia, Bulgaria \\
e-mail: jura@haemimont.bg
\and
NASA\,Goddard Space Flight Center, Code 681, Greenbelt,
MD 20771, USA \\
e-mail: catelan@stars.gsfc.nasa.gov, sweigart@bach.gsfc.nasa.gov
\and
Osservatorio Astronomico di Bologna,
via Zamboni 33, I-10126 Bologna, Italy \\
e-mail: ferraro@astbo3.bo.astro.it
\and
Osservatorio Astronomico di Roma,
I-00040 Monte Porzio Catone, Italy \\
e-mail: buonanno@astrmp.mporzio.astro.it, iannicola@coma.mporzio.astro.it
\and
Sternwarte der Universit\"at Bonn, Auf dem H\"ugel 71, D-53121 Bonn,
Germany\\
e-mail: richtler@astro.uni-bonn.de
}

\date {Received ..... ; accepted .....}

\authorrunning {Borissova et al.}
\titlerunning {Deep photometry for the outer-halo GC NGC~6229}

\maketitle

\begin{abstract}
Deep $BV$ photometry for a large field covering the outer-halo 
globular cluster NGC~6229 is presented. For the first time,
a colour-magnitude diagram (CMD) reaching below the main 
sequence turnoff has been obtained for this cluster. Previous 
results regarding the overall morphology of the horizontal and 
giant branches are confirmed. In addition, nine possible extreme 
horizontal-branch stars have been identified in our deep images, 
as well as thirty-three candidate blue stragglers. We also find 
the latter to be more centrally concentrated than subgiant
branch stars covering the same range in $V$ magnitude.

A comparison of the cluster CMD with the M\,5 (NGC 5904) ridgeline 
from Sandquist et al. (1996) reveals that:

i) NGC~6229 and M\,5 have essentially identical metallicities;

ii) NGC~6229 and M\,5 have the same ages within the errors
in spite of their different horizontal-branch morphologies.

\keywords{Stars: blue stragglers -- Stars: Population II -- Stars:
          Hertzsprung-Russell (HR) diagram -- Galaxy: globular
          clusters: individual: NGC~6229 -- Galaxy: globular
          clusters: individual: M\,5}

\end{abstract}

\section{Introduction}
NGC~6229 (C1645+476) is one of the most remote globular clusters 
(GCs) associated with the Galaxy, lying about 30~kpc from the 
Galactic center (Harris 1996). This outer-halo GC was studied 
photometrically by Cohen (1986) and Carney et al. (1991) and, 
more recently, by Borissova et al. (1997, hereafter BCSS97), 
who presented CCD photometry of the central part of this mildly 
concentrated GC ($c=1.6$). In none of these studies, however, 
was the turnoff (TO) point of the cluster clearly reached, so 
that no conclusions regarding its age could be made.

As discussed extensively in the two previous papers of this 
series (BCSS97; Catelan et al. 1998), NGC~6229 shows both a 
``gap" on the blue horizontal branch (HB) and HB 
bimodality -- unique characteristics among outer-halo GCs. As 
argued by many authors, an understanding of the nature of the 
detected peculiarities in ``bimodal" and ``gap" clusters would 
be of paramount importance for clarifying the nature of the 
so-called second-parameter phenomenon (e.g., Buonanno et al. 
1985; Catelan et al. 1998; Ferraro et al. 1998; Rood et al. 1993; 
Stetson et al. 1996; Sweigart 1999; Sweigart \& Catelan 1998). 
Since age is one of the most prominent second-parameter 
candidates (e.g., Chaboyer et al. 1996b), and since it has 
been suggested that the outer-halo GC system is younger than 
the inner-halo one, obtaining deep photometry below the
turnoff in NGC~6229 is clearly of considerable interest. For 
recent discussions on the ages of outer-halo GCs -- specifically, 
NGC~2419, Palomar~3, Palomar~4, Palomar~14, and Eridanus -- the 
reader is referred to Harris et al. (1997), Sarajedini (1997), 
Stetson et al. (1999), Catelan (1999), and VandenBerg (1999a, 
1999b). 

In this paper we present, for the first time, deep $BV$
photometry reaching below the NGC~6229 TO level. Our 
observational material and data reduction techniques are 
described in Sect.~2 together with our analysis of the 
photometric errors and completeness corrections. Several 
aspects of the colour-magnitude diagram (CMD) of NGC~6229
including candidate blue straggler and extreme HB stars
are described in Sect.~3. The metallicity of NGC~6229,
as determined from various photometric indicators, and its 
age relative to M\,5, as derived from a differential 
application of both the ``vertical" and ``horizontal" 
methods, are then discussed in Sects.~4 and 5, respectively.  
In Sect.~6 we summarize the most important results of the 
present investigation.

In the next paper of this series, a detailed investigation 
of the RR Lyrae population in NGC~6229 will be provided. 

\section{Observations}

\subsection{CCD photometry}
Our analysis is based on the following observational material:

      1. A large set of CCD frames obtained on the nights of 
August 25-26 1987, at the Calar Alto observatory (Spain) of 
the Max-Plank Institute f\"{u}r Astronomie (Heidelberg), 
using the 3.5m telescope equipped with a $1032\times624$ 
RCA chip (hereafter ``CA dataset"). The array scale was
$0.25\arcsec$ ${\rm pixel}^{-1}$, giving a field of view of
about $4.3\arcmin \times 2.6\arcmin$. Only the nine frames
with the best seeing (FWHM better than $1\arcsec$) were 
used to obtain the deep photometry down to $V = 24$~mag 
presented in this paper. Three partially overlapping fields, 
covering a total area of $\sim 22\,{\rm arcmin}^2$ around 
the cluster center, were observed during the run. However, 
only short exposures ($\sim 10\,{\rm sec}$) were obtained 
in the field centered on the cluster center, while long 
exposures (ranging from $\sim 1000\,{\rm sec}$ up to 
$1300\,{\rm sec}$) were secured only in the two more 
external fields placed at $\sim 2\arcmin$ North-West and 
South-West with respect to the cluster center.

      2. Fifty-four $B,V$ frames obtained on June 28-30 
1997, plus six $B,V$ frames obtained on October 6-8 1997 
at the 2m Ritchey-Chr\'etien telescope of the Bulgarian 
National Astronomical Observatory ``Rozhen" with a 
Photometrics $1024\times 1024$ CCD camera (hereafter 
``Rozhen dataset"). The seeing during these observations 
was $\approx 1\arcsec$ with stable and very good photometric 
conditions. The scale at the Cassegrain-focus CCD was 
$0.33\arcsec\,{\rm pixel}^{-1}$ and the observing area 
was $5.6\arcmin \times\ 5.6\arcmin$, centered on the
cluster center. Ten $B$ and 10 $V$ frames obtained under 
the best photometric conditions were added up, resulting 
in total exposure times of $1260\,{\rm sec}$ and 
$1200\,{\rm sec}$, respectively. Two standard fields of 
M\,67 (Christian et al. 1985) were taken before and after 
the observations, for calibration purposes.

\begin{figure}[ht]
      \resizebox{\hsize}{!}{\includegraphics{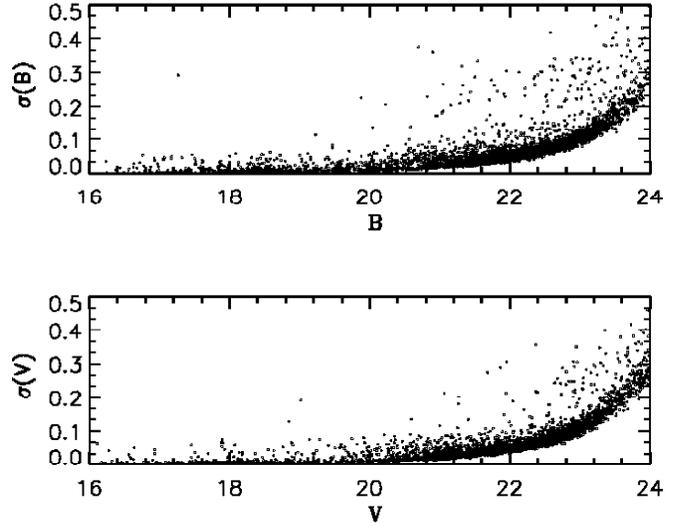}}
      \caption{Formal errors of the Rozhen photometry from 
      the {\sc DAOPHOT} package as a function
      of the mean $B$ and $V$ magnitudes for each star
      }
      \label{Fig01}
\end{figure}

\subsection{Data reduction}

The stellar photometry of the frames was carried out 
independently using {\sc ROMAFOT} (Buonanno et al. 1983) 
for the CA dataset and {\sc DAOPHOT/ALLSTAR} (Stetson 1993) 
for all other frames. Instrumental magnitudes were determined 
for 1500 -- 5000 stars on each frame.

The instrumental values of all the ``Rozhen" frames were
transformed to the standard system using the standard 
field in M\,67 (Christian et al. 1985). Due to unfavorable 
weather conditions during the CA run, no independent 
photometric calibration was possible for the CA dataset. 
For this reason, the magnitudes of this sample were 
originally referred to the $B,V$~Johnson system using a 
large set of stars in common with Carney et al. (1991). 
In order to check the compatibility of the magnitude 
systems of the two datasets, we adopted the following 
procedure: we first transformed the Rozhen $x, y$
coordinate system to the CA local system and then 
searched for stars in common between the two 
datasets -- resulting in a sample of more than 900 stars. 
The residuals in magnitude and colour for the stars in 
common do not show any systematic difference or trend so 
we can conclude that the two datasets are homogeneous in 
magnitude within the errors. This result also suggests that 
the calibration obtained for the Rozhen dataset and adopted 
for the whole dataset presented in this paper nicely agrees 
with the Carney et al. (1991) zero point.

\subsection{Photometric errors}
In order to estimate the internal accuracy of our CCD 
measurements for the Rozhen dataset, we used the formal 
errors from the {\sc DAOPHOT} package and the rms 
frame-to-frame scatter of the instrumental magnitudes. 
The formal errors for all stars vs. their magnitudes are 
displayed in Figure~1. As can be seen, we have large errors
($>0.15$~mag) for $B > 22.5$~mag and $V > 23.0$~mag, which 
therefore define the faint limit of the Rozhen photometry 
($\approx 1.5$~mag fainter than the TO point in $V$).

\begin{figure}[t]
      \resizebox{\hsize}{!}{\includegraphics{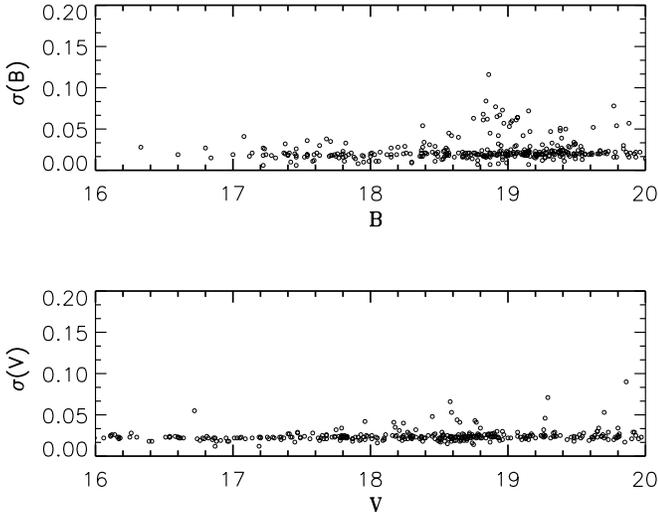}}
      \caption{Rms
      frame-to-frame scatter of the instrumental magnitudes 
      from the Rozhen photometry as a function of the mean 
      $B$ and $V$ magnitudes for each star
      }
      \label{Fig02}
\end{figure}

Following the method described in Ferraro \& Paresce (1993), 
we calculated the rms frame-to-frame scatter of the 
instrumental magnitudes of the 15 $B$ and $V$ frames from 
the Rozhen dataset obtained in the same seeing and FWHM. 
The plot of the rms scatter vs. the mean magnitudes in the 
$B$ and $V$ filters is given in Figure~2. The mean rms 
scatter in both filters over the whole magnitude interval 
is 0.02~mag.

The errors in the CA case, given in Table~1, were determined 
as the rms scatter of multiple measures of stars observed in 
the overlapping regions of adjacent fields. Also given in 
Table~1 are the errors obtained for the Rozhen dataset.  

\begin{table}[h]
\caption {The mean standard deviations of the CA and Rozhen datasets}
\begin{tabular} {ccccc}
\hline
 $V$ range&  $\sigma(B)_{\rm CA}$     &  $\sigma(V)_{\rm CA}$ & 
             $\sigma(B)_{\rm Rozhen}$ &  $\sigma(V)_{\rm Rozhen}$  \\ 
\hline

        $14.0 - 20.0$ &       0.02    &   0.02  &  0.02  &  0.01 \\
        $20.0 - 22.0$ &       0.06    &   0.06  &  0.04  &  0.03 \\
        $22.0 - 24.5$ &       0.10    &   0.10  &  0.12  &  0.15 \\
\hline
\end{tabular}
\label{Tab01}
\end{table}

The final photometric accuracy for the Rozhen and CA datasets should 
include all of these error estimates. From the errors given in 
Figures~1 and 2 and Table~1 we conclude that the total uncertainty is
$\approx 0.03$~mag for $V < 21$~mag and $B < 21$~mag and between
0.05~mag and 0.1~mag for the fainter magnitudes.

\begin{table}[h]
\caption {Completeness functions for the Rozhen dataset}
\begin{tabular} {ccc}
\hline
 $V$ range &  $F(B)$ &   $F(V)$  \\
\hline

        $16.5 - 17.0$ &  1.00 &  1.00   \\
        $17.0 - 17.5$ &  1.00 &  1.00   \\
        $17.5 - 18.0$ &  1.00 &  1.00   \\
        $18.0 - 18.5$ &  1.00 &  1.00   \\
        $18.5 - 19.0$ &  1.00 &  1.00   \\
        $19.0 - 19.5$ &  1.00 &  1.00   \\
        $19.5 - 20.0$ &  1.00 &  1.00   \\
        $20.0 - 20.5$ &  0.99 &  1.00   \\
        $20.5 - 21.0$ &  0.96 &  0.97   \\
        $21.0 - 21.5$ &  0.92 &  0.94   \\
        $21.5 - 22.0$ &  0.86 &  0.88   \\
        $22.0 - 22.5$ &  0.69 &  0.74   \\
        $22.5 - 23.0$ &  0.55 &  0.61   \\
        $23.0 - 23.5$ &  0.38 &  0.44   \\
        $23.5 - 24.0$ &  0.18 &  0.24   \\
        $24.0 - 24.5$ &  0.06 &  0.10   \\
\hline
\end{tabular}
\label{Tab02}
\end{table}

\subsection{Completeness corrections}

The last step in our data reduction was to determine the completeness
functions in the $B$ and $V$ filters. We used the artificial star 
technique (Stetson \& Harris 1988; Stetson 1991a, 1991b) to create a 
series of artificial frames by means of the {\sc ADDSTAR} routine in 
{\sc DAOPHOT~II}. These artificial frames were
then re-reduced in the same manner as the original frames to obtain 
the completeness
functions $F(B)$ and $F(V)$, defined as the ratio of recovered to 
added artificial
stars. The completeness functions for the Rozhen dataset are listed
in Table~2.

\begin{table}[h]
\caption {Number counts for the CA and Rozhen datasets 
for various evolutionary branches} 
\begin{tabular} {lrrc}
\hline
 Branch & CA sample& Rozhen sample& $\frac{N({\rm CA})}{N({\rm Rozhen})}$ \\
\hline
MS           &234     &225     & 1.04     \\
SGB          &236     &278     & 0.85     \\
RGB          &    31  &    26  & 1.19     \\
Red HB       &    6   &    7   & 0.86     \\
RR Lyraes    & 4      &5       & 0.80     \\
Blue HB      &    14  &    10  & 1.40     \\
AGB          &    6   &    6   & 1.00     \\
BSS          &8       &7       & 1.14     \\
Field stars  & 13     &14      & 0.93     \\
\hline
\end{tabular}
\label{Tab03}
\end{table}

Since no specific artificial star test has been performed on the CA 
sample, we can compare the relative completeness degree of the two 
samples from a direct comparison of the star counts found in each 
evolutionary branch. In doing this we selected an area common to 
both datasets relatively far from the cluster center 
($1.0\arcmin < r < 1.5\arcmin$). Then we derived the CMDs over 
this area for the two samples and counted the number of
stars in each branch (with $V<22$~mag) uncorrected for completeness.
The number of stars for the cluster branches and the ratio between 
the numbers found in the CA and Rozhen datasets are given in 
Table~3. As can be seen, the two samples have very closely the 
same overall degree of completeness. 

\begin{figure}[t]
      \resizebox{\hsize}{!}{\includegraphics{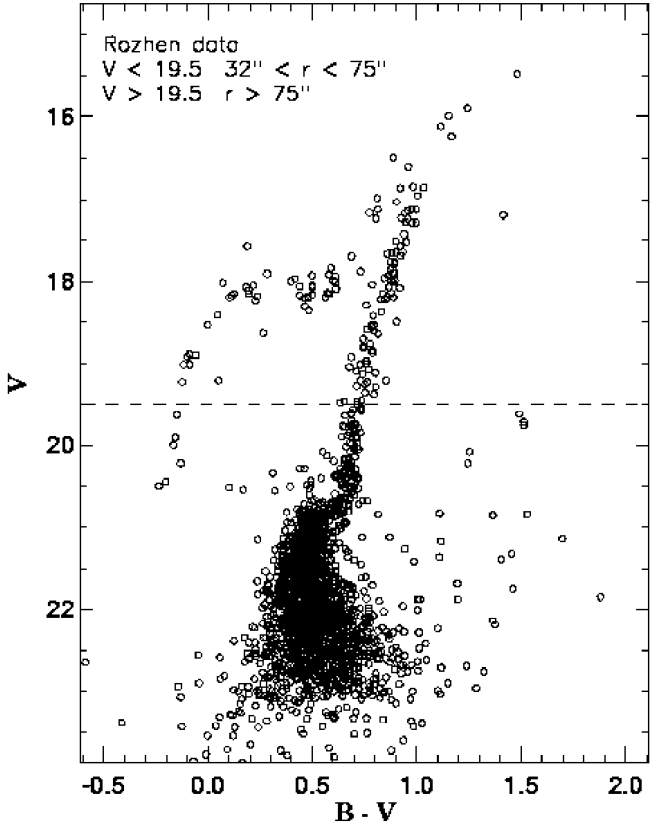}}
      \caption{The $V$, $B-V$ CMD (Rozhen dataset).
      Known variable stars have been omitted from the plot.
      The dashed line at $V=19.5$~mag separates the ``faint" 
      and ``bright" samples, which refer to different areas: 
      $r>75\arcsec$ and $32\arcsec < r < 75\arcsec$, 
      respectively
      }
      \label{Fig03}
\end{figure}

\begin{figure}[t]
      \resizebox{\hsize}{!}{\includegraphics{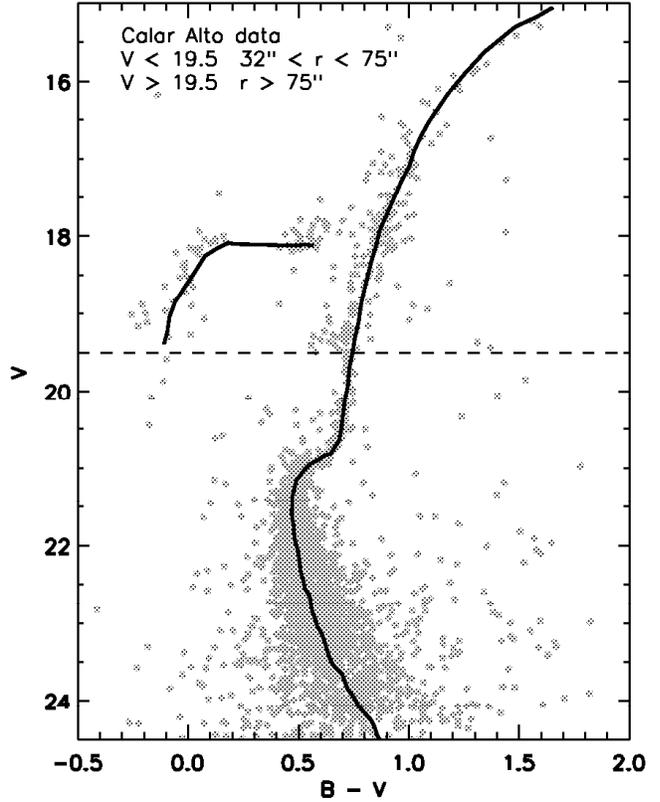}}
      \caption{The $V$, $B-V$ CMD (CA dataset).
      Known variable stars have been omitted from the plot.
      The dashed line at $V=19.5$~mag separates the ``faint" 
      and ``bright" samples, which refer to different areas: 
      $r>75\arcsec$ and $32\arcsec < r < 75\arcsec$, 
      respectively. The M\,5 ridgeline from Sandquist et al. 
      (1996) is overplotted
      }
      \label{Fig04}
\end{figure}

\section{The Colour-Magnitude Diagram}

\subsection{Overall CMD morphology}

The $V$, $B-V$ CMDs for the Rozhen and CA datasets are 
presented in Figures~3 and 4, respectively. The following 
selection criteria were adopted in order to have the best 
definitions of the CMD branches: for $V > 19.5$~mag, only 
stars with $r > 75\arcsec$ from the cluster center were 
plotted [we find the subgiant branch (SGB)-TO region to be 
better defined in this region]; for stars with $V < 19.5$~mag, 
the annular region with $32\arcsec < r < 75\arcsec$ was 
employed (thus avoiding the crowding in the innermost regions 
while minimizing field star contamination). In Figure~4 the 
M\,5 (NGC~5904) ridgeline from Sandquist et al. (1996) is 
overplotted. Vertical and horizontal shifts of 
$\delta\,V = 3.0$~mag and 
$\delta\,(B-V) = 0.0$~mag, respectively, have been applied 
to the M\,5 mean ridgelines. The magnitude shift was 
determined in order to have a good match to the red HB 
of NGC~6229, while the colour shift was derived by matching 
the lower RGBs ($V \sim 20$~mag; see Figure~4 and Figure~9a). 

The $V$, $B-V$ CMD for 818 stars in common between the Rozhen
and CA datasets is presented in Figure~5. To avoid crowding 
in the innermost regions only stars with $r > 32\arcsec$ and 
$V<22$~mag have been plotted. Blue straggler stars (BSS) and 
extreme HB (EHB) candidates (see Sects. 3.3 and 3.4) are 
plotted as asterisks and filled squares, respectively. From 
the comparison of Figure~5 with Figures~3 and 4 it is evident 
that most of these stars are located in the innermost regions
($32\arcsec <r<75\arcsec$).

\begin{figure}[ht]
      \resizebox{\hsize}{!}{\includegraphics{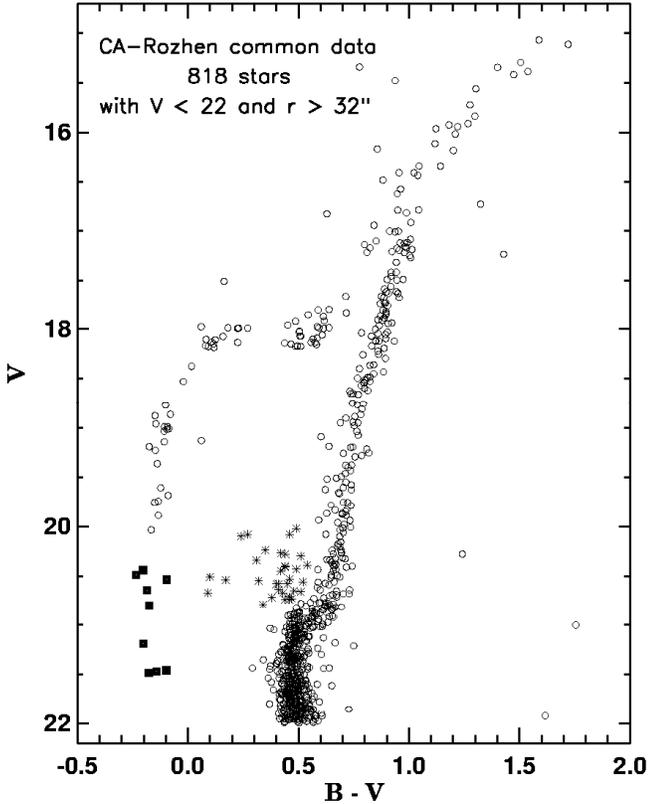}}
      \caption{The $V$, $B-V$ CMD for stars in common
      between the Rozhen and CA datasets with $V < 22$~mag and 
      $r>32\arcsec$ (see text). Filled squares denote candidate 
      extreme horizontal-branch stars; asterisks indicate probable 
      blue straggler stars. Known variable stars are omitted from 
      the plot
      }
      \label{Fig05}
\end{figure}

\begin{figure}[ht]
      \resizebox{\hsize}{!}{\includegraphics{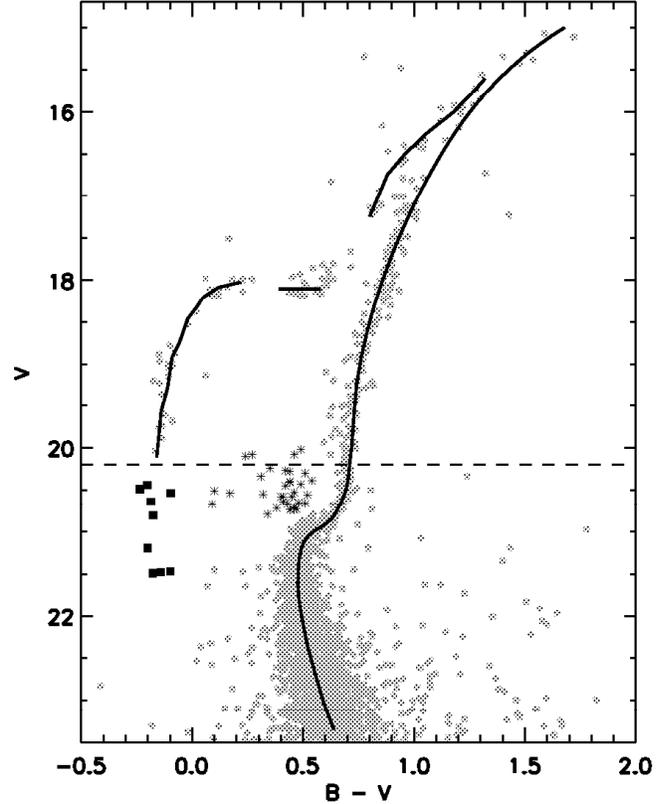}}
      \caption{Detailed plot showing the BSS (asterisks)
      and EHB (filled squares) candidates in NGC~6229. The
      fiducial lines of the cluster are overplotted upon 
      the data in common between the two datasets for 
      $V < 20.2$~mag, and from the CA dataset only 
      ($r > 350$~pixels) for $V > 20.2$~mag. The data 
      in this figure (grey circles) are 
      plotted for illustrative purposes only, and are 
      not representative of the actual CMD of NGC~6229. 
      Note the difference in brightness between the red 
      HB and the tip of the blue HB
      }
      \label{Fig06}
\end{figure}

Our new CMD generally confirms the previous photometric
results of BCSS97. In particular, we can clearly see
the presence of both a red giant branch (RGB) and an
asymptotic giant branch (AGB) population, besides the bimodal
HB with a long blue tail and a ``gap" at $V \simeq 18.4$~mag 
extensively discussed by BCSS97 and 
Catelan et al. (1998). There is a hint of a {\em second} gap,
located at $V \simeq 20$~mag, as previously noted by Carney
et al. (1991). Importantly, however, we have for the
first time clearly identified the main-sequence (MS) TO point
as well as a population of BSS
only hinted at in previous photometric results (see Figure~3
in Carney et al. 1991, and Figure~2 in BCSS97) and a possible
extension of the blue-HB tail to the
EHB region. Each of these new features will be discussed
in the subsections below.

The CMD in Figure~5, augmented by data from the CA dataset 
for $V > 22$~mag, was used to determine the 
fiducial lines of the main branches of NGC~6229
by means of a least-squares fit. The best fit of the upper 
part of the RGB ($V < 19$~mag) was obtained with an
equation of the form:

\begin{equation}
V = a + b\,x + c\,x\,{\rm ln}\,x + d\,x^{0.5}\,{\rm ln}\,x +
 e\,({\rm ln}\,x)^2,
\label{eq1}
\end{equation}

\noindent where $x = B-V$. In
determining the free parameters in eq.~(1) we rejected
stars lying far from the mean branches and suspected field
stars during the several iterations. The std. error of the 
resulting fit is 0.09~mag. The mean ridgelines of the fainter 
RGB ($V > 19$~mag), SGB, MS and BHB were determined by 
dividing these branches into bins and computing in each 
bin the mode of the distribution in colour. Only
the CA dataset was used to define the fiducial
line of the MS region. Ten stars in common with the Rozhen 
dataset for which the {\sc DAOPHOT} parameters {\sc SHARP} 
and {\sc CHI} indicated possible blends were discarded.  
Table~4 presents the adopted
normal points for each branch.  The resulting ridgeline
is plotted in Figure~6, along with the stars from both the 
Rozhen and CA datasets employed to obtain such a ridgeline. 
The 10 possible blends noted above have not been removed
from the plot; they all have $V \sim 20.9$~mag, 
$B-V \sim 0.5$~mag, and can be recognized as a ``clumpy" 
structure immediately below the BSS region in the CMD. 

The MS TO point is found to be at 
$V_{\rm TO} = 21.60\pm0.10$~mag and 
$(B-V)_{\rm TO} = 0.48\pm0.05$~mag.
The quoted errors are the errors of the fitting procedure. 
The magnitude level of the HB, determined as the lower 
boundary of the red HB ``clump", is 
$V_{\rm HB} = 18.10\pm0.05$~mag. This implies a magnitude
difference between the HB and the TO of
$\Delta\,V^{\rm HB}_{\rm TO} = 3.50\pm0.11$~mag.

\subsection{Number ratios}

\begin{table}[t]
\caption {Mean fiducial lines of NGC 6229}
\begin{tabular} {lllllr}
\hline
$V$& $B-V$&    $V$&  $B-V$ & $V$ & $B-V$     \\
\hline
\multicolumn{2}{l}{MS+SGB+RGB} &           &        &  \multicolumn{2}{l}{Red 
HB} \\
15.00    &    1.680   &     18.70 & 0.781     &  18.10      &     0.58  \\
15.10    &    1.620   &     18.80 & 0.772     &  18.10      &     0.48  \\
15.20    &    1.566   &     18.90 & 0.764     &  18.10      &     0.39  \\
15.30    &    1.515   &     19.00 & 0.757     &  \multicolumn{2}{l}{Blue HB}\\
15.40    &    1.469   &     19.10 & 0.750     &  18.02      &     0.22  \\
15.50    &    1.426   &     19.20 & 0.744     &  18.08      &     0.12  \\
15.60    &    1.386   &     19.28 & 0.740    &   18.13      &     0.09  \\
15.70    &    1.349   &     19.43 & 0.735    &  18.20      &     0.05  \\
15.80    &    1.315   &     19.70 & 0.727    &  18.32      &     0.02  \\
15.90    &    1.283   &     19.96 & 0.720    &  18.46      &    $-0.02$  \\
16.00    &    1.253   &     20.21 & 0.709    &  18.61      &    $-0.04$  \\
16.10    &    1.224   &     20.39 & 0.702    &  18.76      &    $-0.06$  \\
16.20    &    1.197   &     20.52 & 0.691    &  18.91      &    $-0.09$  \\
16.30    &    1.172   &     20.63 & 0.675    &  19.08      &    $-0.10$  \\
16.40    &    1.147   &     20.77 & 0.648    &  19.27      &    $-0.11$  \\
16.50    &    1.124   &     20.84 & 0.630    &  19.57      &    $-0.14$  \\
16.60    &    1.102   &     20.90 & 0.606    &  19.85      &    $-0.15$  \\
16.70    &    1.080   &     20.95 & 0.575    &  20.11       &   $-0.16$  \\
16.80    &    1.059   &     21.00 & 0.544    &   \multicolumn{2}{l}{AGB} \\
16.90    &    1.039   &     21.05 & 0.522    &  15.60      &  1.32     \\
17.00    &    1.020   &     21.13 & 0.502    &  15.75      &  1.27     \\
17.10    &    1.001   &     21.20 & 0.491    &  16.00      &  1.18     \\
17.20    &    0.983   &     21.25 & 0.488    &  16.25      &  1.06     \\
17.30    &    0.966   &     21.31 & 0.482    &  16.50      &  0.96     \\
17.40    &    0.949   &     21.42 & 0.479    &  16.75      &  0.88     \\
17.50    &    0.932   &     21.52 & 0.478    &  17.00      &  0.84     \\
17.60    &    0.916   &     21.60 & 0.477    &  17.25      &  0.80     \\
17.70    &    0.901   &     21.70 & 0.478    &             &           \\
17.80    &    0.886   &     21.80 & 0.482    &             &           \\
17.90    &    0.872   &     21.94 & 0.489    &             &           \\
18.00    &    0.858   &     22.08 & 0.499    &             &           \\
18.10    &    0.845   &     22.23 & 0.510    &             &           \\
18.20    &    0.833   &     22.36 & 0.521    &             &           \\
18.30    &    0.821   &     22.63 & 0.549    &             &           \\
18.40    &    0.810   &     23.02 & 0.592    &             &           \\
18.50    &    0.800   &     23.35 & 0.638    &             &           \\
18.60    &    0.790   &           &          &             &           \\
\hline
\end{tabular}
\label{Tab04}
\end{table}

The conclusions from BCSS97 regarding the 
$R = N_{\rm HB}/N_{\rm RGB}$ ratio (Iben 1968) 
between HB stars and RGB stars brighter than the HB are 
confirmed. Using all the available data 
(BCSS97, CA and Rozhen), we find the following number counts: 
$N_{\rm HB} = 285$
(including 9 candidate EHB stars; cf. Sect. 3.4 below),
$N_{\rm RGB} = 258$, and $N_{\rm AGB} = 57$. It thus follows
that $R=1.10\pm0.10$ ($R=1.07\pm0.10$ excluding the
suspected EHB stars). For $R' = N_{\rm HB}/N_{\rm (RGB+AGB)}$,
the values are $R'=0.90\pm\,0.07$ and $R'=0.88\pm\,0.07$,
respectively. Note that the BCSS97 number counts have
been corrected for completeness as described therein. We have
assumed a differential bolometric correction between the HB 
and the RGB of 0.15~mag (Buzzoni et al. 1983). 
As emphasized by Carney et al. (1991) and BCSS97, these $R$ 
and $R'$ values for NGC~6229 indicate either a low cluster 
helium abundance ($Y \approx 0.19$) or the presence of a 
substantial, but as yet undetected, population of EHB stars 
(see BCSS97 for more details). 

As noted by BCSS97, a low helium abundance for NGC~6229 
would lead to a {\em red} HB morphology, as opposed to what 
is observed (Figure~5). In fact, there is even a hint of an 
(extremely hot) EHB population in NGC~6229 (see Sect.~3.4). 

Since in Sect.~5 the age of NGC 6229 will be compared with 
that of M\,5, it is important to investigate any possible 
differences in the chemical composition which could invalidate 
the comparison.\footnote{Note that 
the ``horizontal" method of relative-age determination is 
essentially helium abundance-independent (e.g., Bolte 1990). 
The ``vertical" method of age derivation is also weekly 
dependent on $Y$, provided the adopted GC distance scale 
is {\em model-independent} -- coming, for instance, from 
the Baade-Wesselink method applied to RR Lyrae variables. 
It is only in those instances where the helium abundance  
is consistently used to estimate the HB (and TO) luminosity 
by means of evolutionary models that the ``vertical" method 
becomes substantially dependent on $Y$. This topic is 
discussed at some length in Sect.~4 of Catelan \& de Freitas 
Pacheco (1996).} While the discussion of the metal content 
of the two clusters is presented in Sect.~4, we here compare 
the $R$-ratios in NGC~6229 and M\,5. In doing this comparison 
we shall adopt (in order to be fully consistent with Sandquist 
et al. 1996) their differential bolometric correction between
the HB and the RGB, which turns out to be 0.265~mag for the 
$V$ filter. In this case, we have $N_{\rm RGB} = 275$ and 
the ratios: 
$R = 1.04\pm0.10$ ($R = 1.00\pm0.10$), 
$R'= 0.86\pm0.07$ ($R'= 0.83\pm0.07$). 
These ratios are fully consistent with those obtained by
Sandquist et al. for M\,5 ($R = 1.12\pm0.10$, 
$R' = 0.94\pm0.09$ -- see their Table~7). From this comparison, 
it follows that there is no significant difference in $R$ or 
$R'$ between the two clusters, and that the difference in $Y$ 
between them is unlikely to be higher than $\delta Y \sim 0.01$. 

The consistency of the $R$-ratios derived above would not 
exclude, however, phenomena such as helium mixing amongst 
NGC~6229 red giants, which as pointed out by Sweigart (1997)
would not affect -- to a {\em first approximation} -- the 
number ratios.

\subsection{Blue straggler stars}

None of the previous photometric investigations of NGC~6229 
(Cohen 1986; Carney et al. 1991; BCSS97) addressed the BSS in 
the cluster. Fusi Pecci et al. (1992), on the basis of Figure~3 
in Carney et al (1991), discussed 13 BSS candidates and accordingly 
put NGC 6229 in their group ``BS3", alongside M\,3 (NGC~5272). 
(For a much more detailed discussion of the special BSS population 
in the latter cluster, see Ferraro et al. 1997.)

In order to identify possible BSS candidates in NGC 6229, we 
followed the guidelines from Fusi Pecci et al. (1992, 1993). 
Accordingly, we checked the status of all stars forming an 
extension of the MS up to about 1.5~mag brighter than the TO 
point in $V$, and also the stars located $\approx 0.1$~mag redder 
than the TO. This yielded the thirty-three possible BSS plotted
in Figures~5 and 6. All these stars were measured in both  
the Rozhen and CA datasets. Careful checks were made to ensure 
that the internal errors of the BSS candidates are the same as 
those of the subgiants at the same level. The {\sc DAOPHOT} 
parameters {\sc SHARP} and {\sc CHI} were checked for each BSS 
candidate, in order to ensure that most are unlikely to be the 
result of spurious photometric blends (e.g., Ferraro et al. 1992). 
Follow-up observations are needed to confirm their evolutionary 
status though.  

\begin{table}[t]
\caption {BSS candidates }
\begin{tabular} {lrrcrc}
\hline
No.&   $x$&   $y$&   $V$&   $B-V$&     Distance ($\arcmin$) \\
\hline
1    &    624.90  &  75.997  &  20.27 &   0.42     &   0.53 \\
2    &    355.73  &  322.37  &  20.79 &   0.34     &   0.64 \\
3    &    388.20  &  101.47  &  20.45 &   0.42     &   0.66 \\
4    &    498.71  &  176.58  &  20.67 &   0.43     &   0.71 \\
5    &    384.89  &  132.75  &  20.56 &   0.52     &   0.75 \\
6    &    516.62  &  192.28  &  20.41 &   0.44     &   0.77 \\
7    &     95.32  &  291.92  &  20.02 &   0.49     &   0.77 \\
8    &     510.12 &   207.09 &  20.58 &   0.45     &   0.83 \\
9    &    583.30  &  211.97  &  20.64 &   0.41     &   0.89 \\
10   &    283.93  &  351.55  &  20.08 &   0.46     &   0.90 \\
11   &    498.31  &  224.24  &  20.72 &   0.38     &   0.90 \\
12   &    609.89  &  206.05  &  20.30 &   0.51     &   0.91 \\
13   &    322.69  &  122.20  &  20.74 &   0.46     &   0.93 \\
14   &    482.62  &  233.90  &  20.58 &   0.41     &   0.95 \\
15   &    428.56  &  224.84  &  20.73 &   0.47     &   0.97 \\
c659 &     276.73 &   389.97 &  20.40 &   0.44     &   0.88 \\
c420 &    752.60  &   66.27  &  20.74 &   0.44     &   1.00 \\
c412 &    421.42  &  245.32  &  20.53 &   0.46     &   1.11 \\
c378 &    239.43  &  196.50  &  20.28 &   0.44     &   1.39 \\
c481 &    834.96  &  131.28  &  20.71 &   0.46     &   1.41 \\
c416 &    266.20  &  244.98  &  20.66 &   0.51     &   1.43 \\
c376 &    202.99  &  362.55  &  20.08 &   0.27     &   1.43 \\
c460 &    882.81  &   49.51  &  20.51 &   0.10     &   1.52 \\
c429 &    225.37  &  299.61  &  20.43 &   0.49     &   1.71 \\
c436 &     93.29  &  156.11  &  20.39 &   0.54     &   1.86 \\
c478 &     79.88  &  257.66  &  20.54 &   0.17     &   2.08 \\
c356 &   1088.71  &   99.30  &  20.34 &   0.31     &   2.39 \\
c329 &   675.01   &  204.75  &  20.10 &   0.24     &   1.10 \\
c358 &   452.31   & $-141.51$&  20.65 &   0.48     &   0.66 \\
c425 &   601.18   &$-402.376$&  20.55 &   0.32     &   0.81 \\
c499 &   454.45   &  312.57  &  20.67 &   0.09     &   0.74 \\
c327 &  345.06    &  241.18  &  20.24 &   0.35     &   1.38 \\
c426 &  283.54    &  336.03  &  20.58 &   0.40     &   0.95 \\
\hline
\end{tabular}
\label{Tab05}
\end{table}

According to the Galaxy model of Ratnatunga \& Bahcall (1985), 
the predicted number of field stars in our fields [2 stars at 
$20 < V({\rm mag}) < 21$ and $B-V < 0.8$~mag] is not sufficient 
to explain this population.  

Data for the BSS candidates in NGC~6229 -- magnitudes, colours and
distance from the cluster center -- are listed in Table~\ref{Tab05}.
Their $x$ and $y$ coordinates are in the CA coordinate system.
Eighteen of these candidates are found in the Carney
et al. (1991) photometry list, although only seven
(labeled c425, c359, c426, c358, c460, c376
and c478 in Table~\ref{Tab05})
were selected by Fusi Pecci et al. (1992, 1993).
We were
unable to identify their remaining candidates, possibly due to the 
large errors reported by Carney et al. (1991). Our BSS candidates 
are shown in Figures~5 and 6.

The implied ``specific frequency" of BSS in NGC~6229 
(at $r > 32\arcsec$) is

\begin{equation}
F_{\rm HB}^{\rm BSS} = \frac{N_{\rm BSS}}{N_{\rm HB}} \simeq 0.19.
\label{eq2}
\end{equation}

\noindent This should be compared with the overall frequencies 
recently obtained with HST for a series of GCs: $0.17$ for M\,13 
(NGC~6205); $0.67$ for M\,3; and $\sim 1$ for M\,80 (NGC~6093) 
(Ferraro et al. 1999b). This
would suggest a similar specific frequency of BSS between
NGC~6229 and M13. Note that both M\,13 and NGC~6229 are not 
dynamically evolved (e.g., Trager et al. 1995), as opposed 
to what is likely to be the case for M\,80 -- whose core is 
indeed trying to collapse and stellar interactions may be 
preventing this from happening. The extremely large specific 
frequency of BSS in M\,80 is likely to be due to this effect 
(Ferraro et al. 1999b).  

\begin{figure}[ht]
      \resizebox{\hsize}{!}{\includegraphics{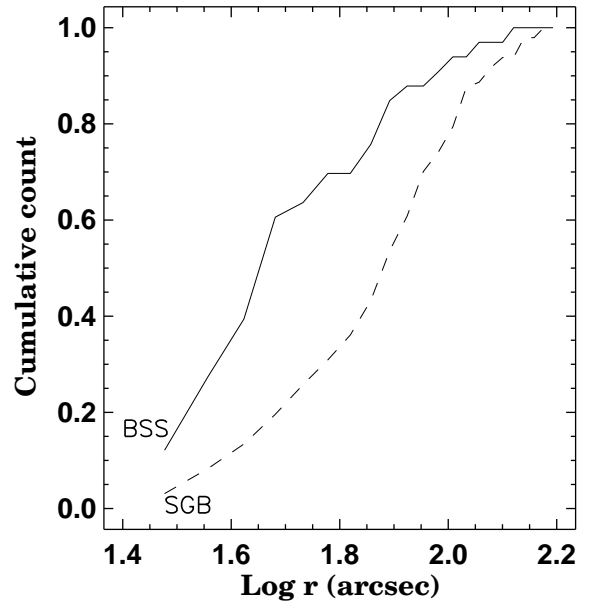}}
      \caption{Radial cumulative distributions of the candidate
      BSS (solid line) and SGB stars (dashed line) in NGC~6229
      }
      \label{Fig07}
\end{figure}

We mention, however, several caveats in comparing the above
HST $F_{\rm HB}^{\rm BSS}$ values with our ground-based value 
for NGC~6229 (see also
Ferraro et al. 1995). In particular, the samples employed to
obtain the NGC~6229 and the HST specific frequencies are:

a) Heterogeneous in sampling: the HST surveys are obtained in
the very central region of each cluster, whereas the NGC~6229
$F_{\rm HB}^{\rm BSS}$ value has been derived for its external
region ($r > 32\arcsec$);

b) Heterogeneous in photometric filters and accuracy: the BSS
samples in HST surveys are based on ultraviolet filters which 
are much more efficient than $B$ and $V$ (as used here) for 
obtaining complete BSS samples;

c) Heterogeneous in defining the ``lower edge" of the BSS
sample: due to the different filter systems, it is unclear
whether the (somewhat arbitrary) lower edges coincide in the 
HST ultraviolet and ground-based $B, V$ studies, possibly 
introducing an important systematic bias.

\noindent We thus caution that the above comparison among
HST $F_{\rm HB}^{\rm BSS}$ values and NGC~6229's might be 
substantially modified once high-accuracy HST data in the 
ultraviolet are available for NGC~6229.

Systematic differences in the radial distribution of BSS with 
respect to SGB stars spanning the same magnitude range have been 
detected in many GCs (I.~Ferraro et al. 1995). The radial 
distribution of the 33~BSS listed in Table~5 was compared with 
the one for 97~SGB stars lying within $\pm 0.1$~mag
in $B-V$ from the cluster fiducial line over the
magnitude range $20 < V({\rm mag}) < 21$. The cumulative radial
distributions for both samples are plotted in Figure~7 as a 
function of the projected distance from the cluster center 
($r > 32\arcsec$). It is evident from this plot that the 
candidate BSS (solid line) are more centrally concentrated 
than the SGB stars (dashed line). A Kolmogorov-Smirnov test 
applied to the two distributions shows that the possible BSS 
population in NGC~6229 is more centrally concentrated than the 
SGB population in the same magnitude interval with $96\,\%$
confidence.

\begin{figure}[ht]
      \resizebox{\hsize}{!}{\includegraphics{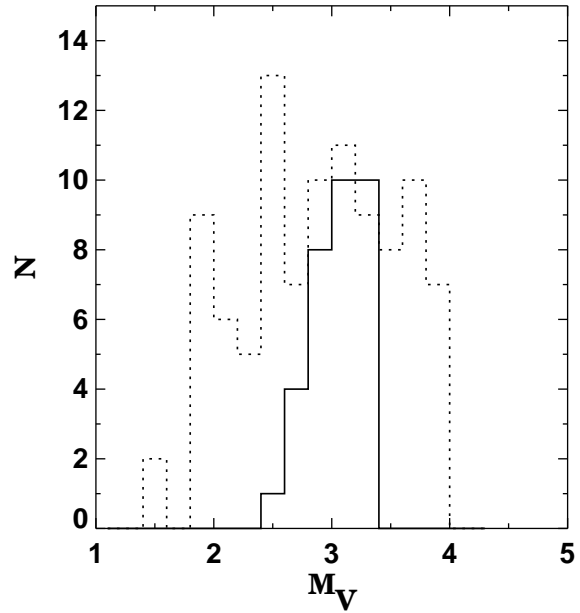}}
      \caption{Luminosity function of the candidate BSS in NGC~6229
      (solid line), compared with that for seven GCs with  
      $\rho_0 > 10^3\,M_{\sun}\,{\rm pc}^{-3}$ from Fusi Pecci 
      et al. (1993) (dashed line)}
      \label{Fig08}
\end{figure}

Figure 8 presents the 
NGC~6229 BSS luminosity function (LF).
This LF seems consistent with the total 
LF presented by Fusi Pecci et al. (1992, 1993) for 625 BSS 
in 25 GCs, displaying a peak at $M_V \simeq 3.2$~mag. It is 
not possible to tell, on the basis of our data alone, whether 
NGC~6229's BSS LF is fully compatible with the ``global" LF 
for GCs with $\rho_0 > 10^3\,M_{\sun}\,{\rm pc}^{-3}$, which 
appears to differ from the one for GCs of lower central 
density (cf. Figure~VIII in Fusi Pecci et al. 1993). 
For comparison purposes, we display in Figure~8 the 
overall LF for BSS in a set of seven GCs with 
$\rho_0 > 10^3\,M_{\sun}\,{\rm pc}^{-3}$ from Fusi Pecci 
et al. (1993) (dashed line), after removing NGC~6229 and 
M\,3 from their sample. For NGC~6229, 
$\rho_0 \simeq 2.5 \times 10^3\,M_{\sun}\,{\rm pc}^{-3}$.   
Note that, for consistency, the distance modulus 
[$(m-M)_V = 17.31$~mag and $E(B-V) = 0.01$~mag] and 
metallicity scales adopted when producing Figure~8 were 
the same as in Fusi Pecci et al. (1993).  

Since NGC~6229 is a crowded, mildly concentrated outer-halo GC,
HST imaging will definitely be necessary in order for a likely
centrally concentrated BSS population to be positively identified
in its innermost regions.

\subsection{Extreme HB stars}

The putative extreme HB (EHB) population suggested by BCSS97 
could not be unquestionably identified in the present study, 
although we found strong hints that it may indeed be present. 
The implications of the presence (or otherwise) of an EHB 
population in NGC~6229 have been discussed at length by BCSS97 
and Catelan et al. (1998).

There are nine stars (marked by filled squares in 
Figures~5 and 6) reaching
as faint as $V = 21.4$~mag, and having $B-V \approx -0.15$~mag,
which resemble an EHB extension to the long blue HB tail of 
NGC~6229. Magnitudes, colours and distances from the cluster 
center for these stars are listed in Table~\ref{Tab06}.
Unfortunately, most of these stars are identified only on the 
CA dataset in the high crowding region and have relatively large 
photometric errors. Two of them, however (No.~5 and c397 in
Table~\ref{Tab06}), were also measured in the Rozhen dataset, 
their photometric parameters being in very good agreement between 
the two datasets. We identify star No.~397 in the Carney et 
al. (1991) photometry list as an EHB candidate. Accurate
photometry extending to even fainter magnitudes is necessary 
to confirm their nature. In particular, in order to identify 
a possible prominent EHB population in NGC~6229 (see BCSS97), 
HST photometry reaching the very crowded cluster core will be
required. We recall that the conspicuous EHB population in 
the core of NGC~2808, a much closer GC ($R_{\sun} = 9.3$~kpc,
compared to $R_{\sun} = 29.3$~kpc for NGC~6229 -- data from
Harris 1996), was {\em only} discovered using HST (Sosin et 
al. 1997), notwithstanding the fact that deep ground-based CCD
photometry had already been published for the outer regions of
NGC~2808 (Ferraro et al. 1990).

\begin{table}[t]
\caption {Extreme HB candidates}
\begin{tabular} {lrrcrc}
\hline
 No.&   $x$&   $y$&   $V$&   $B-V$&     Distance ($\arcmin$) \\
\hline
1 &  562.545 &   86.141 &     20.642 &  $-0.185$&  0.382 \\
2 &  458.270 &  100.294 &     20.535 &  $-0.097$&  0.459 \\
3 &  389.157 &   59.478 &     21.192 &  $-0.201$&  0.572 \\
4 &  458.255 &   141.58 &     20.803 &  $-0.176$&  0.611 \\
5 &  322.760 &  304.384 &     20.490 &  $-0.235$&  0.880 \\
6 &  621.255 &  179.482 &     21.462 &  $-0.098$&  0.839 \\
7 &  473.597 &  223.595 &     21.485 &  $-0.177$&  0.919 \\
8 &  279.580 &  344.230 &     21.471 &  $-0.142$&  0.951 \\
c397 &  675.420 &  309.552 &     20.441 &  $-0.203$&  2.077 \\
\hline
\end{tabular}
\label{Tab06}
\end{table}

\section{Metallicity}

We can derive a {\it photometric} estimate of the cluster 
metallicity from the absolute location in colour of the RGB 
and from its overall morphology. Many observational photometric 
indicators have been proposed and calibrated for this purpose. 
In the following we will use:

$\bullet\,$ $(B-V)_{0,{\rm g}}$,
defined by Sandage \& Smith (1966) as the RGB dereddened colour 
at the HB luminosity level;

$\bullet\,$ $\Delta V_{1.4}$,
defined by Sandage \& Wallerstein (1960) as the magnitude of 
the RGB at $(B-V)_0=1.4$~mag with respect to the HB level. 
Similar parameters $\Delta V_{1.1}$ and $\Delta V_{1.2}$, 
measured at $(B-V)_0=1.1$~mag and $(B-V)_0=1.2$~mag, 
respectively, have been recently defined by Sarajedini \& 
Layden (1997);

$\bullet\,$ $(B-V)_{0,-2}$, defined by Sarajedini \& 
Layden (1997) as the RGB colour at $M_V = -2$~mag;

$\bullet\,$ $S$, defined by Hartwick (1968) as the slope 
of the RGB measured between two points along the RGB: the 
first at $V_{\rm HB}$ and the second at 
$V=V_{\rm HB}-2.5\,{\rm mag}$;

$\bullet\,$ $\Delta V^{\rm HB}_{\rm bump}$, the location of 
the RGB ``bump" in the CMD relative to the HB level. A 
recalibration of this indicator as a function of cluster 
metallicity has been recently provided by Sarajedini \& 
Forrester (1995), expanding on the dataset used in the 
original calibration (Fusi Pecci et al. 1990).

In order to measure these parameters for NGC~6229, we first 
need to adopt values for the interstellar reddening, distance 
modulus and $V_{\rm HB}$. For consistency with the values 
used in the previous sections (except for Sect.~3.2), we 
will adopt:  
$V_{\rm HB}=18.10\pm 0.05$~mag,
$E(B-V)=0.01$~mag and $(m-M)_V=17.37$~mag (Harris 1996).
From these values, the mean ridgeline listed in Table~4, 
and the location of the RGB ``bump" from BCSS97, we derive:
$(B-V)_{0,{\rm g}}=0.84\pm0.03$~mag,
$\Delta V_{1.1} =1.54\pm0.07$~mag,
$\Delta V_{1.2} =1.95\pm0.07$~mag,
$\Delta V_{1.4} =2.56\pm0.07$~mag,
$(B-V)_{0,-2}=1.47\pm0.05$~mag,
$S=4.62\pm0.07$, and
$\Delta V^{\rm HB}_{\rm bump}=-0.05\pm0.08$~mag.

An additional estimate of the cluster metallicity can be 
derived by the so-called ``simultaneous reddening and 
metallicity (SRM) method" (Sarajedini 1994; Sarajedini \& 
Layden 1997). This method was created to derive the reddening 
and metallicity of a GC simultaneously from its RGB and HB 
photometric properties.

From each of these observables we can obtain an independent
{\it photometric} estimate of the cluster metallicity by 
using the corresponding calibration in terms of ${\rm [Fe/H]}$ 
reported in the literature. Most of the available calibrations 
are based on the Zinn \& West (1984, hereafter ZW84) metallicity 
scale, although some recent papers (Carretta \& Bragaglia 1998; 
Ferraro et al. 1999a) have presented recalibrations in terms of 
the new scale by Carretta \& Gratton (1997, hereafter CG97), 
based on direct high-resolution spectroscopy for RGB stars in 
a set of 24 GCs.

Results based on each available metallicity indicator on the two
metallicity scales are summarized in Table~7.

We note that the reddening obtained from the SRM method 
based on the Sarajedini \& Layden (1997) calibration, 
$E(B-V) = 0.02 \pm 0.01$~mag, is fully consistent with the 
values commonly found in the literature (e.g., Harris 1996).

On the {\em Zinn \& West (1984) scale}, the weighted mean
photometrically-inferred NGC~6229 metallicity is:

\begin{equation}
{\rm [Fe/H]}_{\rm ZW84\,\,scale}=-1.4\pm 0.1~{\rm dex}.
\label{eq3}
\end{equation}

Instead, on the {\em Carretta \& Gratton (1997) scale} the
NGC~6229 metallicity is:

\begin{equation}
{\rm [Fe/H]}_{\rm CG97\,\,scale}= -1.1\pm 0.1~{\rm dex}.
\label{eq4}
\end{equation}

Note that the quoted uncertainties in the metallicity estimates 
provided in Table~7 are derived only from the formal propagation 
of the errors in the photometric indicators. However, in the 
final values we assume a conservative global error of 0.1~dex.

It is important to note that a substantially lower 
metallicity (${\rm [M/H]} = -1.4\pm 0.1$) than provided 
in the CG97 scale (eq.~4) has recently been obtained for 
NGC~6229 from direct medium-resolution spectroscopy of 
bright cluster giants (Wachter et al. 1998). 

However, particular care is required in comparing different
metallicty scales, since most of them are found not to be  
mutually consistent. Carretta \& Gratton (1998) noted that
in the Wachter et al. (1998) scale Galactic GCs turn out to 
be systematically more metal poor with respect to the CG97 
scale. In fact, considering two clusters (namely, M\,13 and 
NGC~7006) examined with both the Wachter et al. and CG97 
techniques, they found: 

$$
{\rm [Fe/H]}_{\rm CG97\,\,scale} = 
{\rm [M/H]}_{\rm Wachter\,\,et\,\,al.\,\,scale} + 0.2~{\rm dex}.
$$

Taking into account this offset, the metallicity found by 
Wachter et al. (1998) for NGC~6229 translates into a 
value  

$${\rm [Fe/H]}_{\rm CG97\,\,scale} = -1.2\pm 0.1~{\rm dex}$$

\noindent in the CG97 scale, 
which is fully compatible with the photometric estimate
we found above (see eq.~4). Note that Pilachowski et al.
(1983) derived ${\rm [Fe/H]} \simeq -1.3$ from echelle 
spectroscopy of a red giant (IV-12) in NGC~6229; in 
their scale, however, 47~Tuc (NGC~104) had a surprisingly
low metallicity (${\rm [Fe/H]} = -1.09$), thus indicating 
that the ``true" NGC~6229 metallicity should be higher 
than implied by the ZW84 scale.    

Since the CMDs of NGC~6229 and M\,5 agree remarkably 
well at the level of the RGB, the two clusters must have 
essentially identical (photometric) metallicities. 
This conclusion does not depend upon (possible) deep  
mixing effects among NGC~6229's and/or M\,5's giants, 
since mixing affects only marginally the RGB locus and 
shape (Sweigart 1997). 

Thus we can conclude that both spectroscopic and 
photometric measurements suggest that M\,5's (and 
thus NGC~6229's) metallicity is indeed quite high: 
${\rm [Fe/H]} = -1.1\pm 0.1$ (see also Sneden et al. 
1992).\footnote{It is interesting to 
note that, according to the latest isochrones computed 
by VandenBerg (1998), it is not possible to fit M5's CMD 
{\em unless} ${\rm [Fe/H]} < -1.3$ (even including the 
observed enhancement in the $\alpha$-elements; Sneden et 
al. 1992), thus favoring the ZW84 scale. On the other hand, 
assuming the ``global metallicity" [M/H] and adopting the 
latest models by Chieffi et al. (1998), Ferraro et al. 
(1999a) have been able to remove the disagreement between 
theory and observations in regard to the location of the 
RGB luminosity function ``bump" (e.g., Fusi Pecci et al. 
1990) in a sample of 47 Galactic GCs -- favoring instead 
the CG97 scale. Possible explanations for the discrepancy 
between theory and observations, as far as the ``bump" 
location goes, have been summarized by Catelan \& de 
Freitas Pacheco (1996; see their Sect.~5).}  

\begin{table}[t]
\caption {Metallicity of NGC~6229}
\begin{tabular} {lcl}
\hline
Parameter & [Fe/H] & Reference \\
\hline
\multicolumn{3}{c}{Zinn \& West (1984) metallicity scale} \\
\hline
$(B-V)_{0,{\rm g}}$ & $-1.39 \pm 0.15$ & Zinn \& West (1984)        \\
                    & $-1.40 \pm 0.13$ & Gratton (1987)             \\
                    & $-1.26 \pm 0.13$ & Costar \& Smith (1988)     \\
                    & $-1.37 \pm 0.10$ & Gratton \& Ortolani (1989) \\
                    & $-1.36 \pm 0.15$ & Sarajedini \& Layden (1997)\\
                    & $-1.36 \pm 0.05$ & {\it Weighted mean} \\
$\Delta V_{1.4}$    & $-1.45 \pm 0.07$ & Zinn \& West (1984)        \\
                    & $-1.29 \pm 0.07$ & Costar \& Smith (1988)     \\
                    & $-1.38 \pm 0.05$ & Gratton \& Ortolani (1989) \\
                    & $-1.37 \pm 0.04$ & {\it Weighted mean} \\
$\Delta V_{1.1}$    & $-1.33 \pm 0.06$ & Sarajedini \& Layden (1997)\\
$\Delta V_{1.2}$    & $-1.33 \pm 0.06$ & Sarajedini \& Layden (1997)\\
$(B-V)_{0, -2}$     & $-1.31 \pm 0.05$ & Sarajedini \& Layden (1997)\\
$S$                 & $-1.35 \pm 0.05$ & Gratton \& Ortolani (1989) \\
$\Delta V^{\rm HB}_{\rm bump}$ & $-1.40 \pm 0.10$ & Sarajedini \& Forrester 
(1997) \\
``SRM method"       & $-1.44 \pm 0.08$ & Sarajedini \& Layden (1997)\\
\hline
\multicolumn{3}{c}{Carretta \& Gratton (1997) metallicity scale} \\
\hline
$(B-V)_{0,{\rm g}}$ & $-1.12 \pm 0.08$ & Carretta \& Bragaglia (1998)\\
$\Delta V_{1.1}$    & $-1.06 \pm 0.08$ & Carretta \& Bragaglia (1998)\\
$\Delta V_{1.2}$    & $-1.10 \pm 0.08$ & Carretta \& Bragaglia (1998)\\
\hline
\end{tabular}
\label{Tab07}
\end{table}

\section{Age determination}

As already stated, our photometric material allowed us 
to obtain, for the first time, a CMD for NGC~6229 
reaching below the cluster TO point and suitable for 
an age determination. In Figure~4, the cluster CMD 
obtained from the CA dataset is displayed together 
with the M\,5 ridgeline from Sandquist et al. (1996). 
Note that a comparison between the outer-halo GC 
NGC~6229 and M\,5 appears especially interesting in 
light of the discovery that M\,5's orbital parameters 
imply that it too is an outer-halo GC, which just happens
to be currently close to its perigalacticon (Cudworth 1997 
and references therein).

Comparing the NGC~6229 CMD with the one for M\,5, we reach 
the following conclusions.

Assuming that M\,5 and NGC~6229 have the same metallicity 
(see Sect.~4) and using the remarkable fit of the M\,5 
ridgeline to the NGC~6229 CMD in the SGB-TO region (see 
Figure~4), it immediately follows that M\,5 and NGC~6229 
have very similar ages (VandenBerg et al. 1990). A similar 
conclusion follows from a comparison with the NGC~1851
ridgeline (Walker 1992; see also Figure~7 in Walker 1998). 
Figure~4 shows, however, that the match between the HBs is 
not very good at the red end of the blue HB, which appears
brighter in NGC~6229 than in M\,5. The red-HB loci agree 
quite well, though. (A comparison between the RR Lyrae 
properties is deferred to Paper~IV in our series.) This 
is similar to what has been claimed by Stetson et al. 
(1996) for the bimodal-HB cluster NGC~1851. BCSS97 had 
already hinted that the red end of the blue HB of NGC~6229 
might be brighter than the red HB of the cluster (see also 
Figure~6). Figure~9a shows a more detailed comparison 
between the NGC 6229 and M\,5 CMDs around the TO and HB 
regions. That the blue HB of NGC~6229 is brighter than 
M\,5's is quite evident in this plot. The same is not 
true when a comparison is made between NGC~6229 and 
NGC~1851, however (Figure~9b). This may shed light on the 
nature of the second-parameter effect (Stetson et al. 1996; 
Catelan et al. 1998; Sweigart 1999; Sweigart \& Catelan 
1998; Walker 1998).

\begin{figure*}[htbp]
      \resizebox{\hsize}{!}{\includegraphics{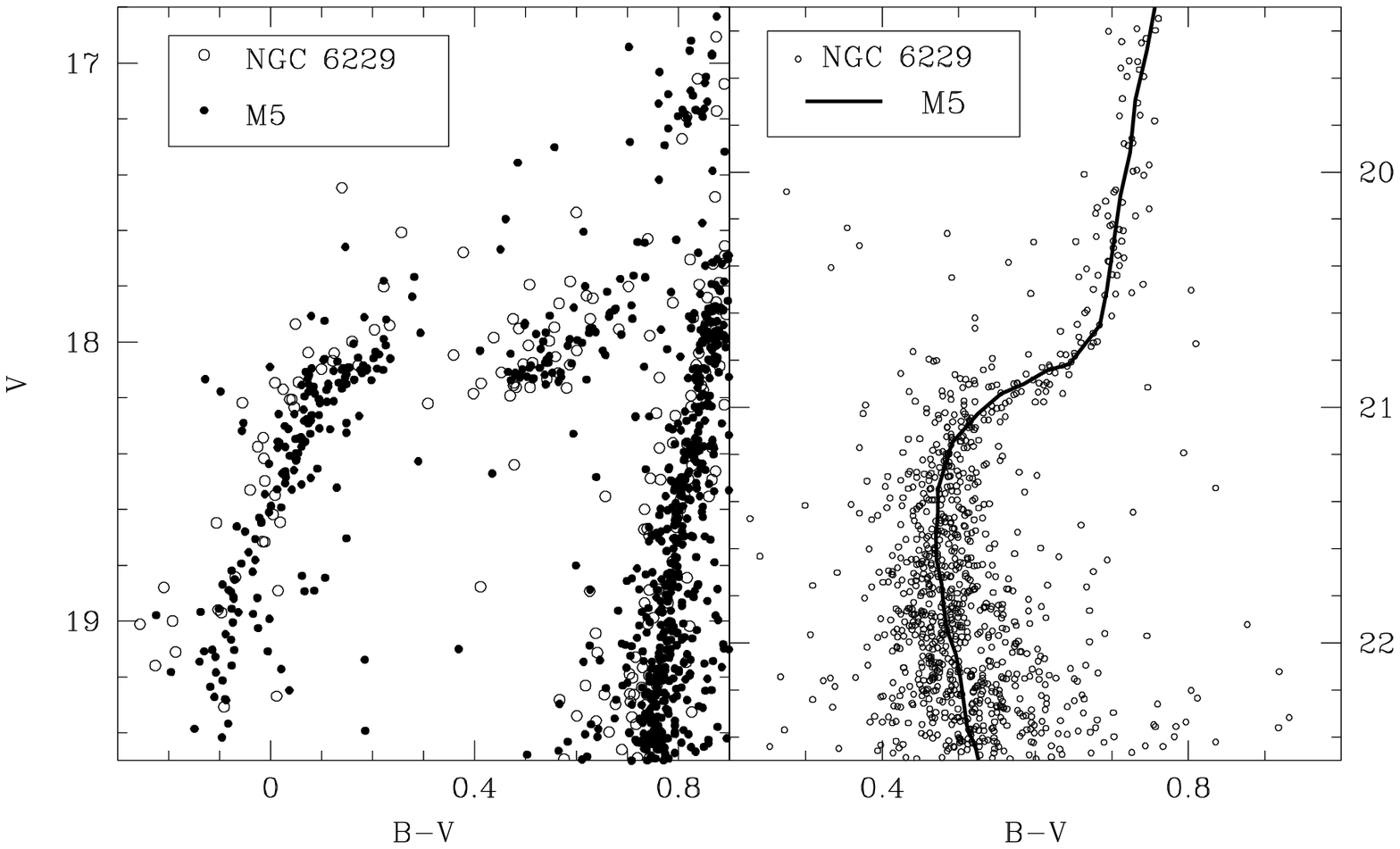}}
      \resizebox{\hsize}{!}{\includegraphics{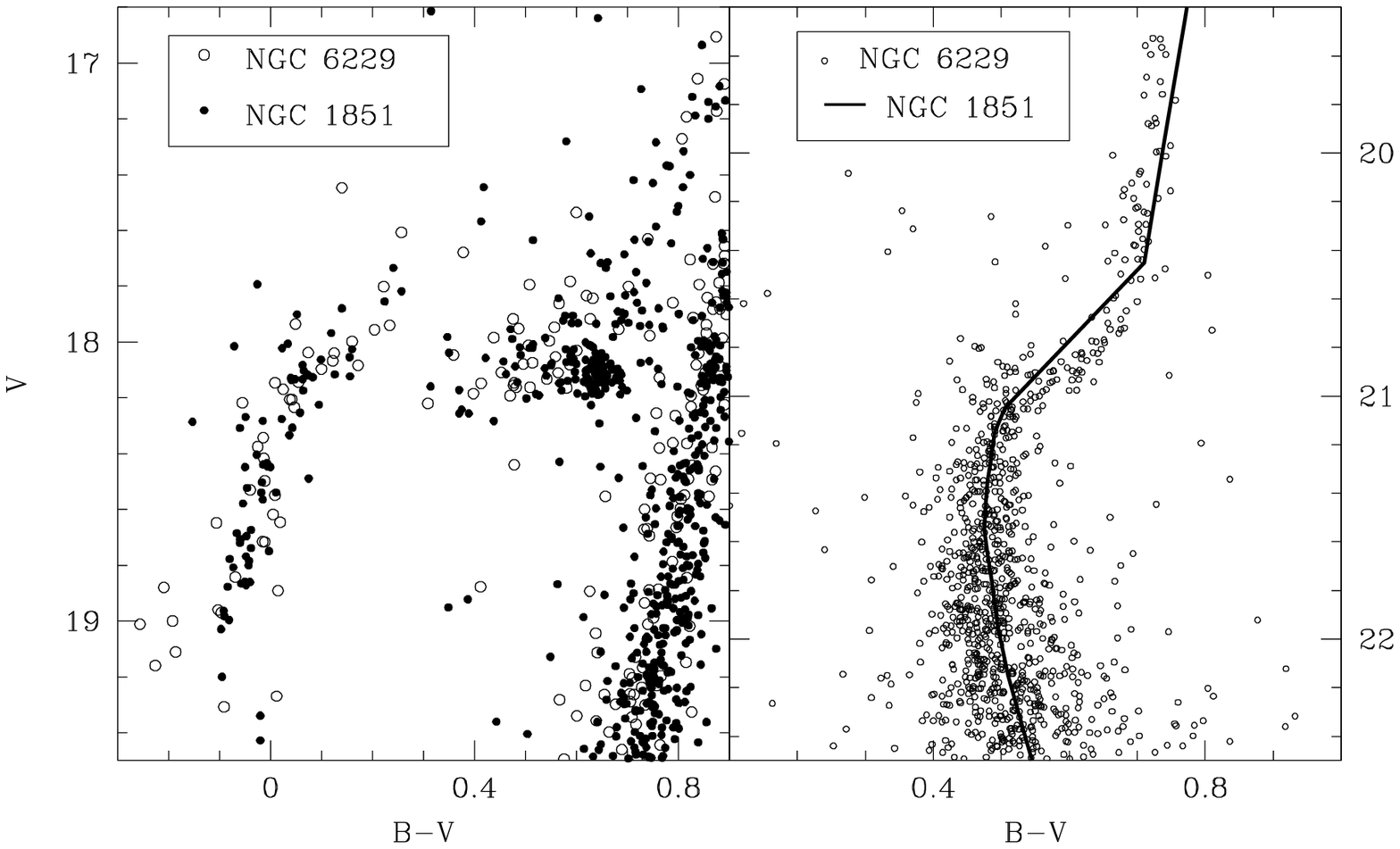}}
      \caption{a) Detailed comparison between the NGC~6229 
       and M\,5 CMDs: in the {\it left-hand panel} the HB 
       region for stars from the CA dataset with $32''<r<75''$ 
       has been plotted. In the {\it right-hand panel}, the 
       SGB-TO region for stars from the CA dataset with $r>75''$ 
       has been plotted. The solid line represents the M\,5 
       ridgeline from Sandquist et al. (1996). b) Same as a),
       except that NGC~1851 has been used for a comparison 
       (data from Walker 1992). Open circles represent NGC~6229 
       and filled circles either M\,5 or NGC~1851. {\em In both 
       panels, known variables have been removed}
      }
      \label{Fig09}
\end{figure*}

To confirm quantitatively that NGC~6229 and M\,5 are 
essentially coeval, we have applied both ``vertical" 
and ``horizontal" methods for determining relative ages 
from the magnitude difference $\Delta V_{\rm TO}^{\rm HB}$ 
between the HB and the TO and the color difference 
$\Delta (B-V)_{\rm TO}^{\rm RGB}$ between the RGB and 
the TO. In order to overcome the problems related to the 
calibration and use of {\em absolute} quantities, we have 
applied these age parameters in a strictly {\em differential} 
sense by referring NGC~6229 to our reference GC M\,5, 
following the approach of Buonanno et al. (1993) and 
I.~Ferraro et al. (1995).

\begin{figure}[htbp]
      \resizebox{\hsize}{!}{\includegraphics{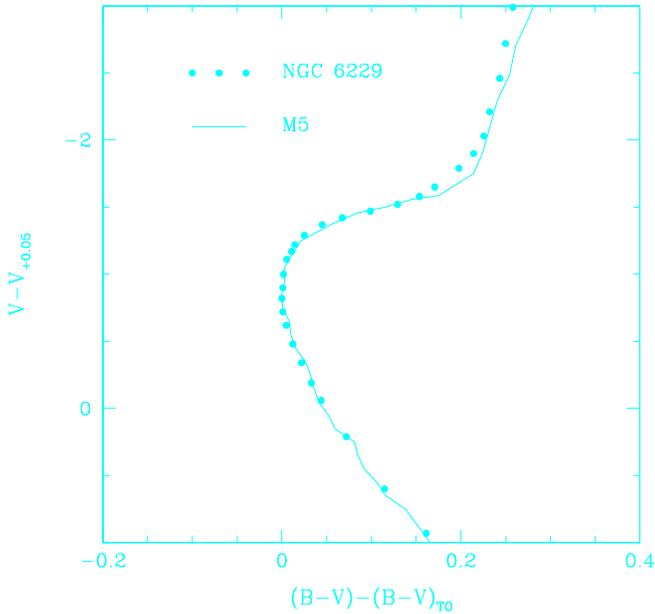}}
      \caption{Comparison between the ridgelines for NGC~6229
      ($\bullet$) and M\,5 (---), registered following the
      VandenBerg et al. (1990) prescriptions in order to obtain
      a measurement of their relative ages
      }
      \label{Fig10}
\end{figure}

On the basis of the VandenBerg \& Bell (1985) models, 
Buonanno et al. (1993) obtained the following ``vertical" 
relation between the differential age parameter
$\Delta = \Delta_1 V_{\rm TO}^{\rm HB}-
                      \Delta_2 V_{\rm TO}^{\rm HB}$
and the age $t_9$ in Gyr:

$$\Delta \log t_9 = (0.44 + 0.04\,{\rm [Fe/H]})\,\Delta,$$
where subscript ``1" stands for NGC~6229, and ``2" for M\,5.
Assuming ${\rm [Fe/H]} \sim -1.1$ for both GCs and using the 
value of $\Delta V_{\rm TO}^{\rm HB}$ derived in Sect.~3.1 
for NGC~6229, we obtained:
$\Delta = 3.50 - 3.47 = 0.03\pm0.13$~mag, implying
$\Delta \log t_9 \sim 0.0119$ and thus

$$\Delta\,t_9 ({\rm NGC~6229} - {\rm M\,5}) \sim 
                               +0.4 \pm 2.0\,\,{\rm Gyr}.$$
\noindent 

The ``horizontal'' method has been applied following the 
procedure described by VandenBerg et al. (1990). In 
Figure~10 we compare the M\,5 and NGC~6229 fiducial lines 
after registering them as recommended by VandenBerg et al. 
According to that paper the colour difference 
$\Delta (B-V)_{\rm TO}^{\rm RGB}$ is computed at the point 
along the RGB where $V-V_{+0.05}=-2.5$~mag. Here $V_{+0.05}$ 
is the MS magnitude at which $B-V$ is 0.05~mag redder than 
the TO. From Figure~10 we derive a value of $-0.010\pm0.015$ 
for the differential age parameter
$\delta = \Delta_1 (B-V)_{\rm TO}^{\rm RGB}-
                          \Delta_2 (B-V)_{\rm TO}^{\rm RGB}$,
defined by Buonanno et al. (1993).  Using the calibration of
$\delta$ given by VandenBerg \& Stetson (1991), namely,

$$\Delta \log t_9 = (-4.32 - 0.78\,{\rm [Fe/H]})\,\delta,$$
we obtain an age difference between NGC~6229 and M\,5 of

$$\Delta\,t_9 ({\rm NGC~6229} - {\rm M\,5}) \sim 
                               +1.2 \pm 1.5\,\,{\rm Gyr}.$$

A new age parameter has been recently defined by Buonanno 
et al. (1998): $\Delta V^{0.05}$, which is the magnitude 
difference between the HB and the point on the upper MS, 
where $B-V$ is 0.05~mag redder than the TO (see Figure~1 in 
Buonanno et al.). For M\,5 and NGC~6229, we find that:
$\Delta V^{0.05}=4.31\pm0.10$~mag and $4.32\pm0.11$~mag, 
respectively. This implies an age difference

$$\Delta\,t_9 ({\rm NGC~6229} - {\rm M\,5}) \sim 
                                 +0.1 \pm 1.5\,\,{\rm Gyr}$$

\noindent between these two GCs.

In summary, all the employed methods agree within the errors
that NGC~6229 and M\,5 are essentially coeval.\footnote{In 
the present paper, we have chosen not to employ the 
recently-proposed relative-age indicator from Chaboyer et 
al. (1996a), which relies on the region of the CMD which is 
brighter than the TO and redder by 0.05~mag in $B-V$. We 
find that this region is particularly subject to spurious 
photometric blends (Sect.~3.1), and thus -- unlike what is 
claimed by Chaboyer et al. -- is subject to potentially 
large observational errors.}

The weighted mean age difference between NGC~6229 and M\,5 
from the three determinations is:

\begin{equation}
\Delta\,t_9 ({\rm NGC~6229} - {\rm M\,5}) \simeq 
                                +0.5\pm 1.0\,\,{\rm Gyr}.
\label{eq5}
\end{equation}

\noindent For a comparison of the age of M\,5 with those of 
other GCs of comparable metallicity, we refer the reader to 
VandenBerg et al. (1990) and VandenBerg (1999b).

\section{Summary and concluding remarks}

In the present paper we have presented the first deep CMD 
for the outer-halo globular cluster NGC~6229 from data 
obtained at the Calar Alto and Rozhen observatories. The 
combined CMD reaches $\sim 2$~mag below the main-sequence 
turnoff of the cluster.

Our detailed analysis of the CMD properties reveals that:

1. The perfect alignment of the upper RGBs in NGC 6229 and
M\,5 suggests that they must have the same metallicity. This 
is confirmed by our {\it photometric} estimate of NGC 6229's 
metallicity  
(${\rm [Fe/H]_{CG97\,\,scale}} = -1.1\pm 0.1$ dex), 
which fully agrees with the spectroscopic measurements 
obtained by CG97 and Sneden et al. (1992) for M\,5
(${\rm [Fe/H]_{CG97\,\,scale}} = -1.1\pm 0.1$ dex, 
$[\alpha/{\rm Fe}] \simeq +0.2$ dex).

Our {\em photometric} estimate of the NGC~6229 metallicity
has been indirectly confirmed by the recent spectroscopic 
measurements carried out by Wachter et al. (1998), who found
${\rm [M/H]} = -1.4\pm 0.1$~dex. Once the offset between the 
Wachter et al. and CG97 metallicity scales found for GCs in
common between the two studies is properly taken into account 
(Carretta \& Gratton 1998), this corresponds to 
${\rm [Fe/H]_{CG97\,\,scale}} = -1.2\pm 0.1$~dex;

2. The CMD of NGC~6229 is entirely compatible with the one for 
M\,5 (Sandquist et al. 1996), implying that the two clusters 
have nearly identical ages. Indeed, a quantitative measurement 
of the age difference between the two GCs using three different 
methods shows them to have the same ages to within 
$\approx 1$~Gyr. However, the HB morphologies of the two 
clusters do differ in detail, NGC~6229's extending to bluer 
colours than M\,5's (with a likely extreme HB population 
present) and showing a ``dip" in the number counts at the 
RR Lyrae level (see the Appendix in Catelan et al. 1998 for 
a discussion);  

3. Thirty-three candidate blue straggler stars have been 
identified;

4. Nine possible extreme HB stars have been detected. If 
confirmed, this will be the first known case of an outer-halo 
GC with an extreme HB population. Associated with this 
population, we find a hint of a (second) ``gap" at the faint 
end of the blue HB. NGC~6229 is the only known outer-halo 
globular with HB gap(s) and a bimodal HB (BCSS97; Catelan 
et al. 1998);

5. Confirmation of the BSS and EHB populations, as well as a
detailed analysis of the radial population gradients discussed 
by BCSS97 and in Sect.~5.2, must await deep HST photometry of 
the crowded central regions of this remote and quite 
concentrated globular cluster.

\begin{acknowledgements}
The authors are grateful to E. Carretta, R. Gratton, F. Grundahl,
D. A. VandenBerg and the referee, B. Carney, for very helpful 
remarks. This research was supported in part by the Bulgarian 
National Science Foundation grant under contract No.~F-604/1996 
with the Bulgarian Ministry of Education and Sciences. 
Support for M.C. was provided by NASA through Hubble Fellowship
grant HF--01105.01--98A awarded by the Space Telescope Science
Institute, which is operated by the Association of Universities
for Research in Astronomy, Inc., for NASA under contract
NAS~5-26555.
\end{acknowledgements}

%

\end{document}